\documentclass{nature}
\usepackage{amsmath}
\usepackage[dvipsnames]{xcolor}
\usepackage[euler]{textgreek}
\usepackage{hyperref}
\usepackage{multirow}

\usepackage[labelfont=bf]{caption}
\usepackage[version=4]{mhchem}
\usepackage[noabbrev,capitalize]{cleveref}
\usepackage{mathptmx}
\usepackage{pdfpages}
\newcommand{\psquared}{$\mathcal{P}^2$}

\title{\psquared: Combining pressure and electrochemistry to synthesize superhydrides}

\author{Pin-Wen Guan$^1$, Russell J. Hemley$^{2,3}$,  Venkatasubramanian Viswanathan$^{1,4}$}

\usepackage{graphicx}
\makeatletter
\let\saved@includegraphics\includegraphics
\AtBeginDocument{\let\includegraphics\saved@includegraphics}
\makeatother
\graphicspath{ {./fig/} }

\begin{document}

\maketitle

\begin{affiliations}
 \item Department of Mechanical Engineering, Carnegie Mellon University, Pittsburgh, Pennsylvania 15213, USA
 \item Department of Physics, University of Illinois at Chicago, Chicago, IL 60607 USA
 \item Department of Chemistry, University of Illinois at Chicago, Chicago, IL 60607 USA
 \item Department of Physics, Carnegie Mellon University, Pittsburgh, Pennsylvania 15213, USA
\end{affiliations}

\newpage
% Please give the surname of the lead author for the running footer

\begin{abstract}
Recently, superhydrides have been computationally identified and subsequently synthesized with a variety of metals at very high pressures.  In this work, we evaluate the possibility of synthesizing superhydrides by uniquely combining electrochemistry and applied pressure.  We perform computational searches for palladium superhydrides using density functional theory and particle swarm optimization calculations over a broad range of pressures and electrode potentials. We incorporate exchange-correlation functional uncertainty using the Bayesian error estimation formalism to quantify the uncertainty associated with the identified stable phases.  Based on a thermodynamic analysis, we construct pressure-potential phase diagrams and provide an alternate synthesis concept, \psquared~ (pressure-potential), to accessing novel phases having high hydrogen content. Palladium-hydrogen is a widely-studied material system with the highest hydride phase being \ce{Pd3H4}. Most strikingly for this system, at potentials above hydrogen evolution and $\sim$300 MPa pressure, we find the possibility to make palladium superhydrides (e.g., \ce{PdH10}).  We demonstrate the generalizability of this approach for La-H, Y-H and Mg-H with 10-100 fold reduction in required pressure for stabilizing phases.  In addition, the \psquared~ strategy allows stabilizing new phases that cannot be done purely by either pressure or potential and is a general approach that is likely to work for synthesizing other superhydrides at modest pressures.  
\end{abstract}

Metal hydrides produced under pressure containing large amounts of hydrogen are currently of great interest. Most notable are the superhydrides (defined as MH$_n$, for $n>6$) first predicted\cite{liu2017potential,peng2017hydrogen}, synthesized\cite{geballe2018synthesis} and discovered\cite{PhysRevLett.122.027001} to be near room temperature superconductors in the La-H system. The Pd-H system has long been specifically important in both fundamental science and technological applications\cite{flanagan1991palladium,viswanathan1998metal} for potential applications in superconductivity\cite{skoskiewicz1972superconductivity}, hydrogen uptake\cite{boudart1975solubility}, and low-energy nuclear reactions\cite{berlinguette2019revisiting}. Experimentally, the highest hydride phase synthesized is \ce{Pd3H4}\cite{fukai1994formation}, under high pressure conditions.  

Here we explore the phase stability of selected metal hydrides, possibly superhydride phases, as a function of pressure and electrochemical conditions (electrode potential, pH) using density-functional theory (DFT)–based structure search methods. We also examine the dependence of the results on the choice of exchange correlation functional within the Bayesian error estimation formalism. We begin with a detailed consideration of the important Pd-H system. After a detailed discussion of the identified stable structures, we present a calculated pressure-dependent electrochemical phase diagram for Pd-H. Based on these calculations, we identify an alternate approach, which we term \psquared, to access Pd-H phases having high hydrogen content.  Most strikingly, we find that it is thermodynamically feasible to electrochemically synthesize \ce{PdH10} at a modest pressure of 300 MPa (3 kbar).  The stability line of phase boundary between PdH and \ce{PdH10} shows that three orders of magnitude in pressure could be compensated by a modest electrochemical driving force of $\sim$0.1 V.  We show generalizability of the approach to La, Y and Mg hydrides, which opens new opportunities for the creation of metal superhydrides and other novel materials by combining pressure and electrochemical loading techniques.

%\section{Results and discussion}
\section*{Structure search and analysis}

The phase stability of metal hydrides over a wide range of compressions was explored using density-functional theory combined with the CALYPSO structure search method\cite{Wang2010,Wang2012} to identify the range of stable structures and stoichiometries possible. Uncertainty quantification becomes important given the challenges in describing high pressure phases\cite{Amsler2018}.  The Bayesian error estimation functional with van der Waals correlation (BEEF-vdW) was employed to provide a confidence value (c-value) for competing phases to avoid possible bias due to selection of a particular DFT functional\cite{houchins}, an approach that has been successfully applied to calculate uncertainty in phase diagrams for other systems\cite{Guan2019,Houchins2020}. To enhance robust assessment of the ground state within the structure search, we used the ensemble of functionals within the BEEF formulation to identify the predicted ground state. Thus, each functional identifies a particular ground state for a given composition and c-value quantifies what fraction of the functionals identify that structure as having the minimum energy.

\begin{figure}
\includegraphics[width=0.49\linewidth]{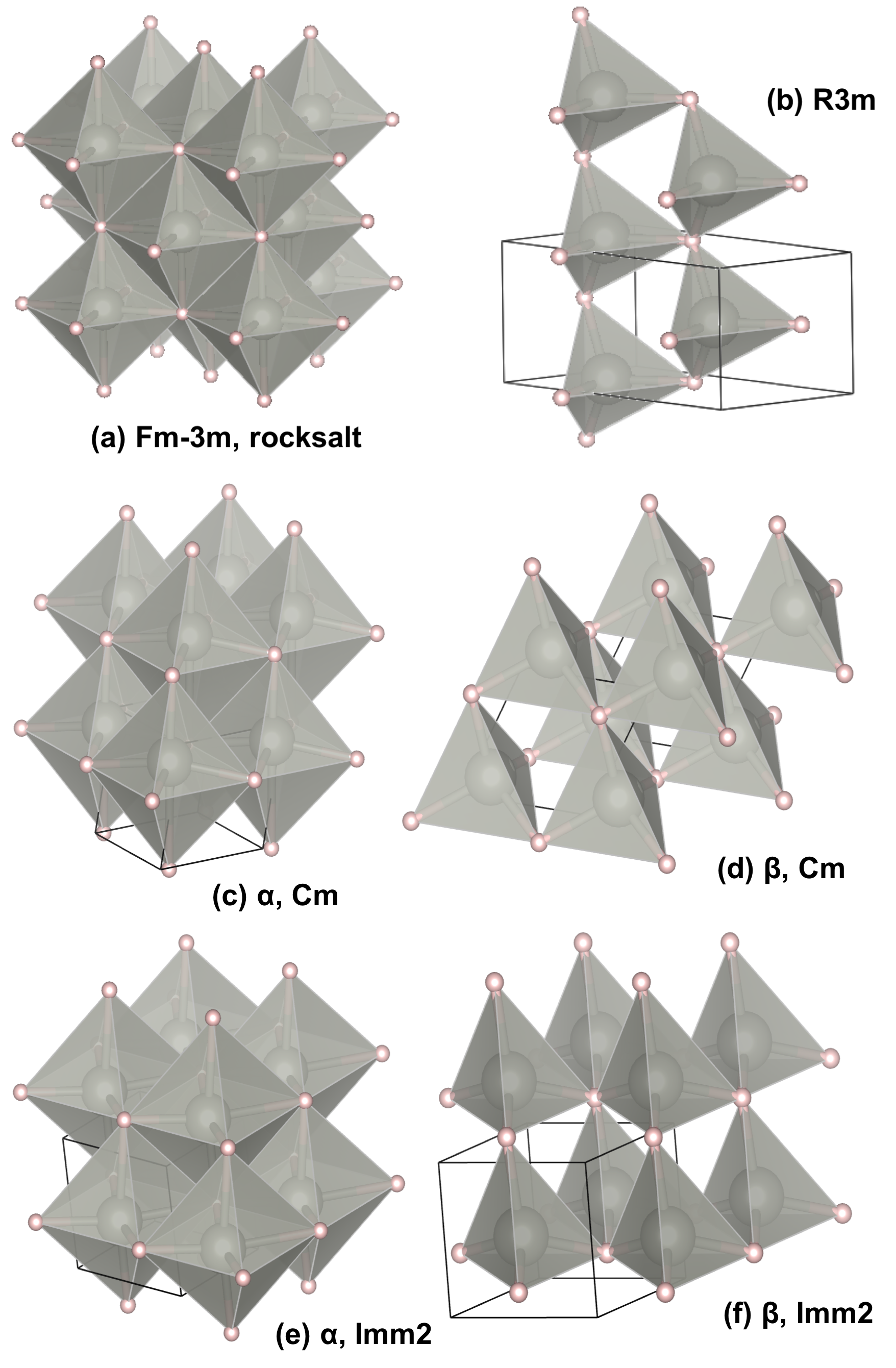}
\includegraphics[width=0.49\linewidth]{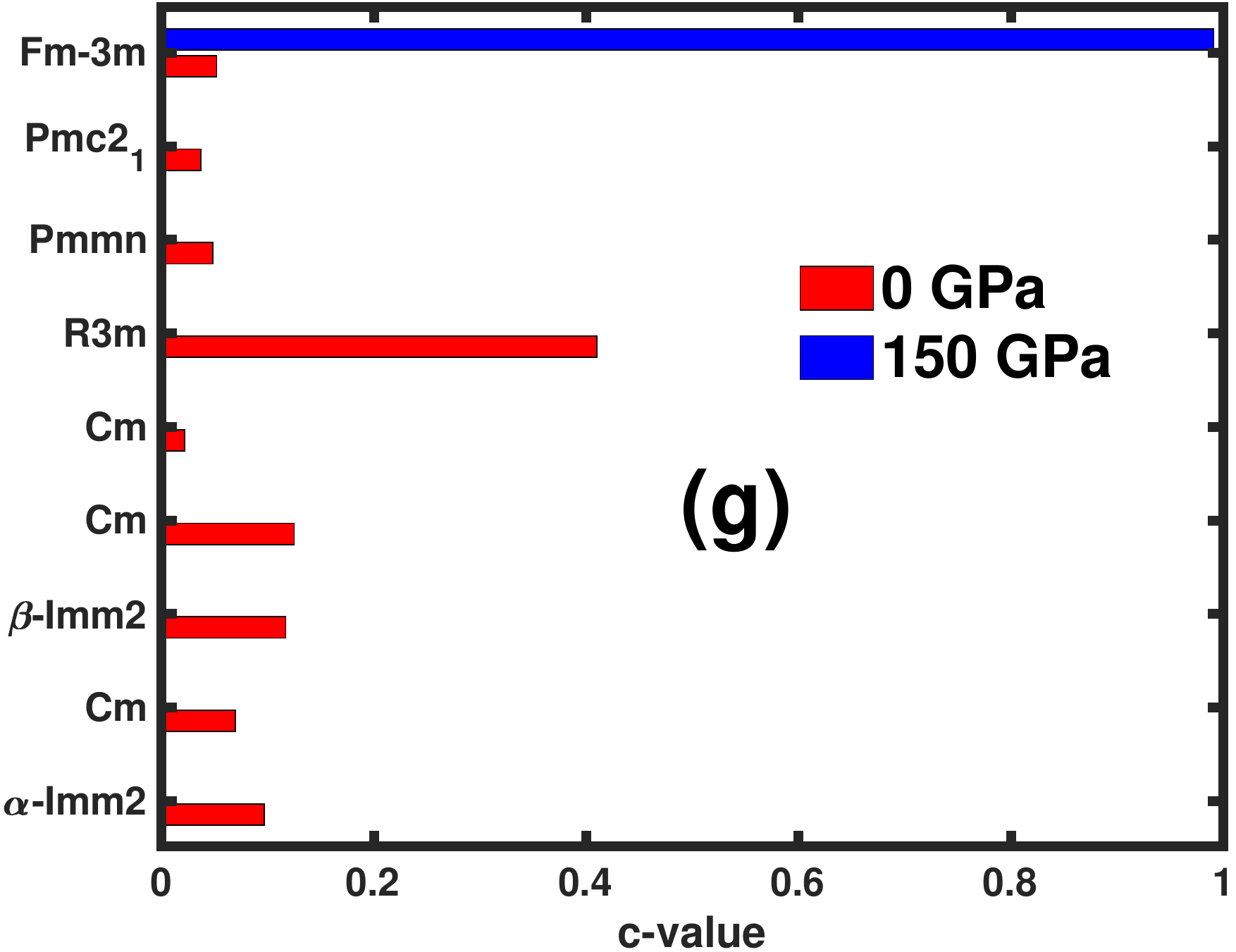}
\centering
\caption{Crystal structures of PdH (a) $Fm\Bar{3}m$ and (b) $R3m$. In the $Fm\Bar{3}m$ structure, the Pd atom is octahedrally coordinated whereas in $R3m$, Pd is tetrahedrally coordinated. (c) Ground states with c-values of PdH at zero pressure and 150 GPa calculated by BEEF-vdW. The most probable structure has the $R3m$ space group at 0 GPa although several other structures are also possible. At 150 GPa, the $Fm\Bar{3}m$ structure is the most probable.}
\label{fig:c-PdH}
\end{figure}

We begin by discussing the results for Pd-H at zero pressure and megabar (150 GPa) pressures in the absence of electrochemical loading. Two experimentally assessed compositions at ambient and high pressure, PdH and \ce{Pd3H4}, were considered first.  We find several structures close in energy for PdH, and used the Bayesian error estimation capability within BEEF-vdw to construct a confidence-value (c-value) diagram (Fig. \ref{fig:c-PdH}). This analysis indicates that the most probable predicted ground state of PdH has the $R3m$ space group with Pd in tetrahedral coordination.  However, this structure has a c-value of around 0.4, indicating that at GGA-level DFT, it is not possible to conclusively identify the true ground state.  We recover the experimentally reported $Fm\Bar{3}m$ rocksalt structure as one of the possible structures, though it has a lower c-value (about 0.05).  At high pressure (e.g., 150 GPa), the method gives the rocksalt structure as the most probable with a high c-value, indicating that on compression this structure is indeed the ground state predicted at the GGA-level of theory. These results are consistent with experiments reported up to 100 GPa\cite{brownsberger2017x,guigue2020x}.  

We next consider \ce{Pd3H4}, which is the only higher hydride reported experimentally, e.g., at around 5 GPa\cite{fukai1994formation}. The structure search calculations identify a structure with Pd having 5-fold and 6-fold coordination and space group $Cm$ as the most probable (Fig. S1).  We recover the experimentally observed \ce{Cu3Au}-type structure with the $Pm\Bar{3}m$ space group as one of the probable structures.  This structure can be viewed as introducing one Pd vacancy in each unit cell of rocksalt PdH, and therefore can also be written \ce{Pd3VaH4}, where Va represents a vacancy.  At high pressure, we find a low-symmetry $P1$ structure with a complex coordination environment as the most probable structure.

\begin{figure}
\includegraphics[width=0.5\linewidth]{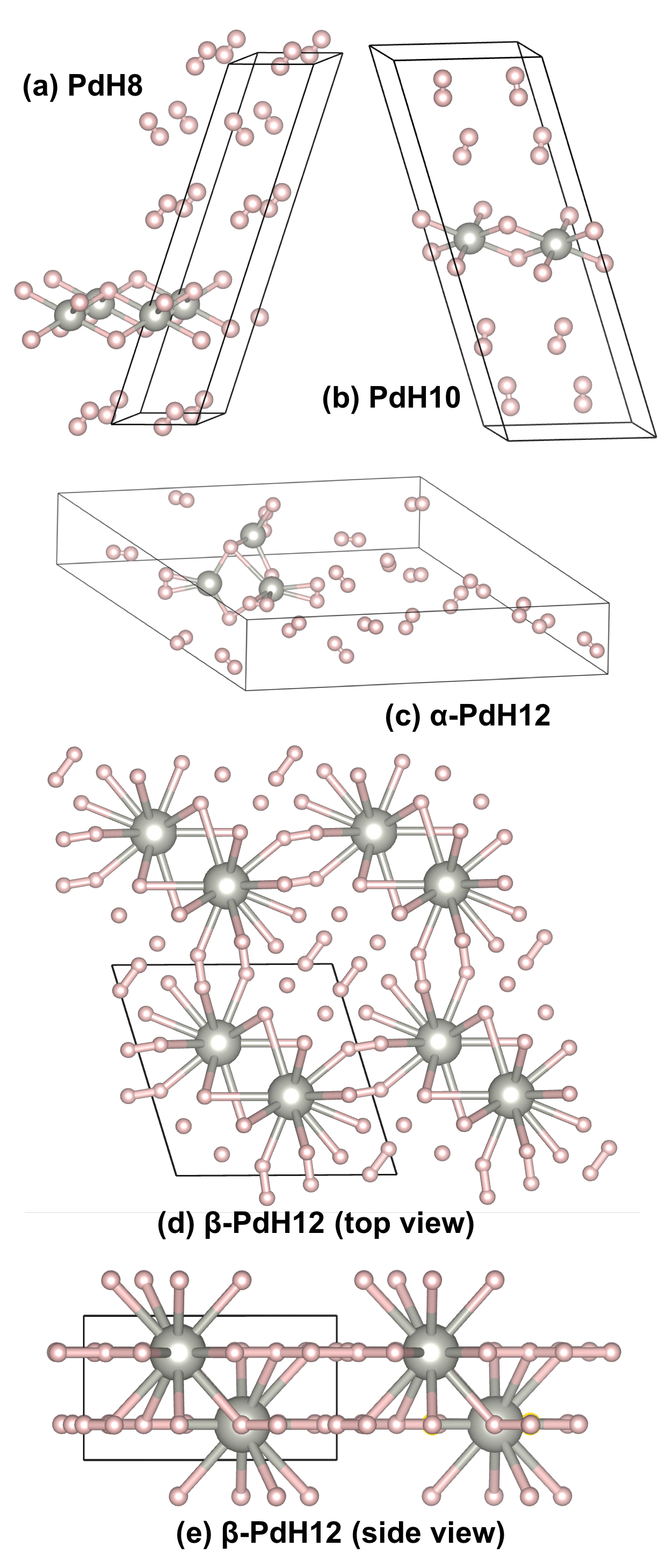}
\centering
\caption{Results of ground-state structure search for (a) \ce{PdH8}, (b) \ce{PdH10}, (c) \ce{PdH12} at zero pressure and \ce{PdH12} (d) top and (e) side view at 150 GPa.}
\label{fig:0&150-PdHn}
\end{figure}

Following this analysis, we examine the stable structures of compositions PdH$_n$ where n is an integer between 2 and 12. The most probable structures identified in the structure search are shown in Fig. \ref{fig:0&150-PdHn}.  These structures consist of Pd-H layers or clusters between which \ce{H2} molecules are located. The atomic coordination features of the low enthalpy structures are characterized by the average radial distribution function (RDF) (Figs. S9 and S10). All structures at zero pressure are characterized by Pd-H and H-H bonding but variable bond distances. Whereas nearest neighbor Pd-H distances remain within 1.7-2.0 {\AA}, the nearest neighbor H-H distances change significantly with composition. For \ce{PdH_n} when n is less than 2-3, the H-H distances are mainly over 2 {\AA}; however, when n is larger than 2-3, H-H forms a peak at about 0.7-0.8 {\AA}, indicating the formation of \ce{H2} molecules. Pressure also affects the RDF significantly. At 150 GPa, for example, H-H distances span a broad range beyond the conventional covalent \ce{H2} bond length, indicating diverse coordination environments of H at high compressions.

\begin{figure}
\includegraphics[width=0.7\linewidth]{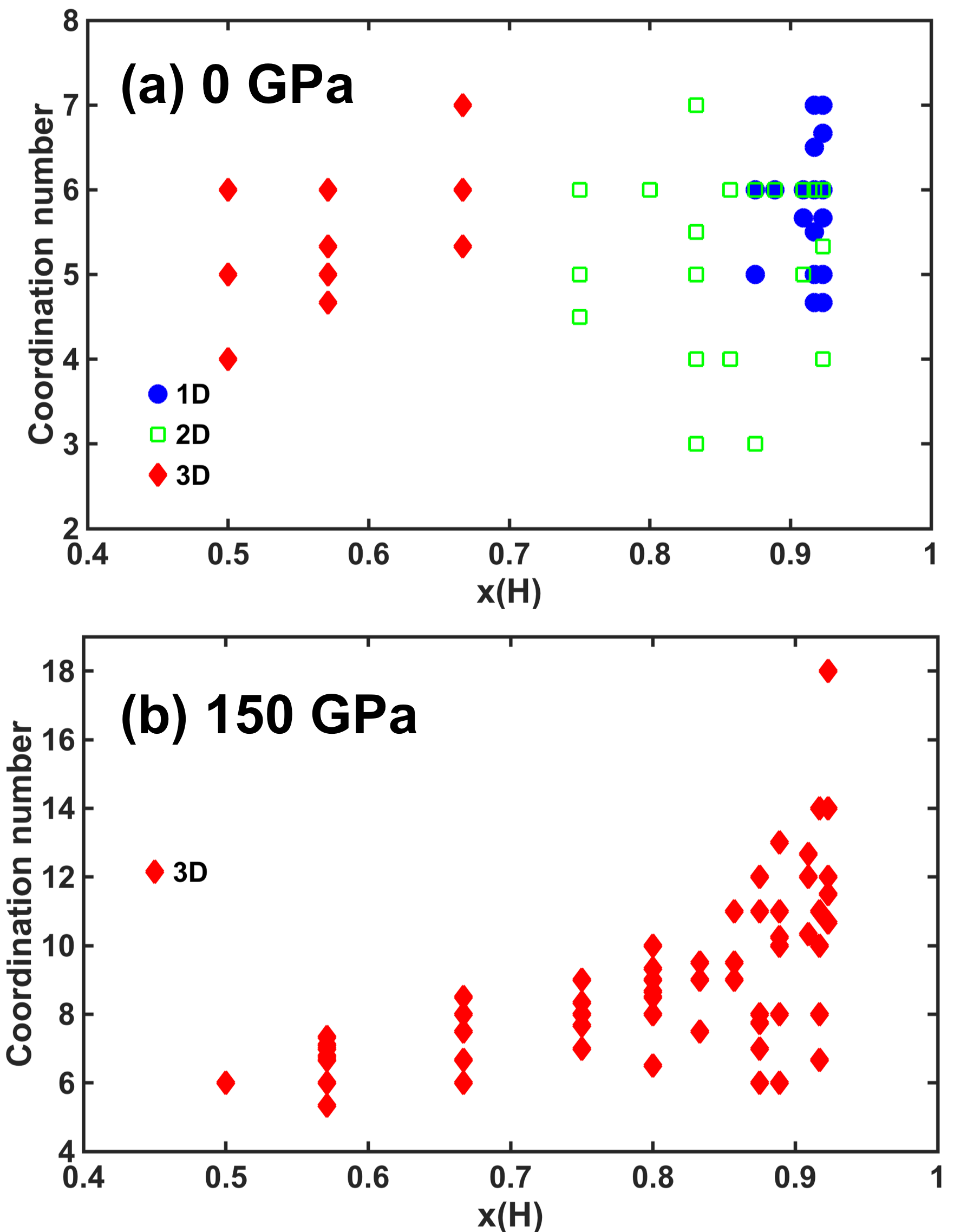}
\centering
\caption{Composition-dependent topological features of the low enthalpy structures at (a) zero pressure and (b) 150 GPa. The coordination number corresponds to the number of nearest neighbor H atoms surrounding a Pd atom, and the dimensionality is used to describe the framework formed by the Pd polyhedra.}
\label{fig:topo}
\end{figure}

To further characterize the structural features of these Pd hydrides, a topological analysis of the low enthalpy structures was performed (Fig. \ref{fig:topo}). Each structure is based on a Pd-H framework with the void space occupied by extra H atoms if any, which are mainly present in the form of \ce{H2} molecules. The basic structural unit in the Pd-H framework is a polyhedron with the Pd atom coordinated by H atoms. The Pd-centered polyhedra can be directly connected with each in terms of sharing of faces, edges or corners, or indirectly connected via intermediate H atoms, forming a network with specific dimensionality. In a 0D framework, the Pd-centered polyhedra are isolated, whereas they can also form columns and layers to give 1D and 2D frameworks, respectively. The network can also extend in three dimensions (3D). Obviously, both pressure and composition have significant influence on the coordination number (CN) of Pd and the dimensionality of the Pd-H framework. At zero pressure, the CN is between 3 and 7, and moderately increases from PdH to \ce{PdH2} but saturates beyond \ce{PdH2}, with the maximum CN pinned at 7. The dimensionality also changes around \ce{PdH2} and the transition is even sharper. From PdH to \ce{PdH2}, the frameworks are exclusively 3D, whereas beyond \ce{PdH2}, no 3D framework is present in the structures examined. From \ce{PdH3} to \ce{PdH6}, the frameworks are exclusively 2D, while starting from \ce{PdH7}, 1D gradually becomes dominant. At 150 GPa, the frameworks are invariably 3D regardless of the compositions, but CN is diffuse, spanning between 5-18, and shows a dramatic change with composition. Based on the above analysis, it can be concluded that high pressure can compress large amounts of H around Pd and increase the connectivity of the framework components. 
%Furthermore, since it is already shown that any Pd superhydride under the studied pressure is thermodynamically unstable against PdH and \ce{H2}, high pressure seems necessity in that it can generate the 3D framework with possibility to keep lattice integrity after H2 molecules escape from the framework.      

\begin{figure}
\includegraphics[width=0.45\linewidth]{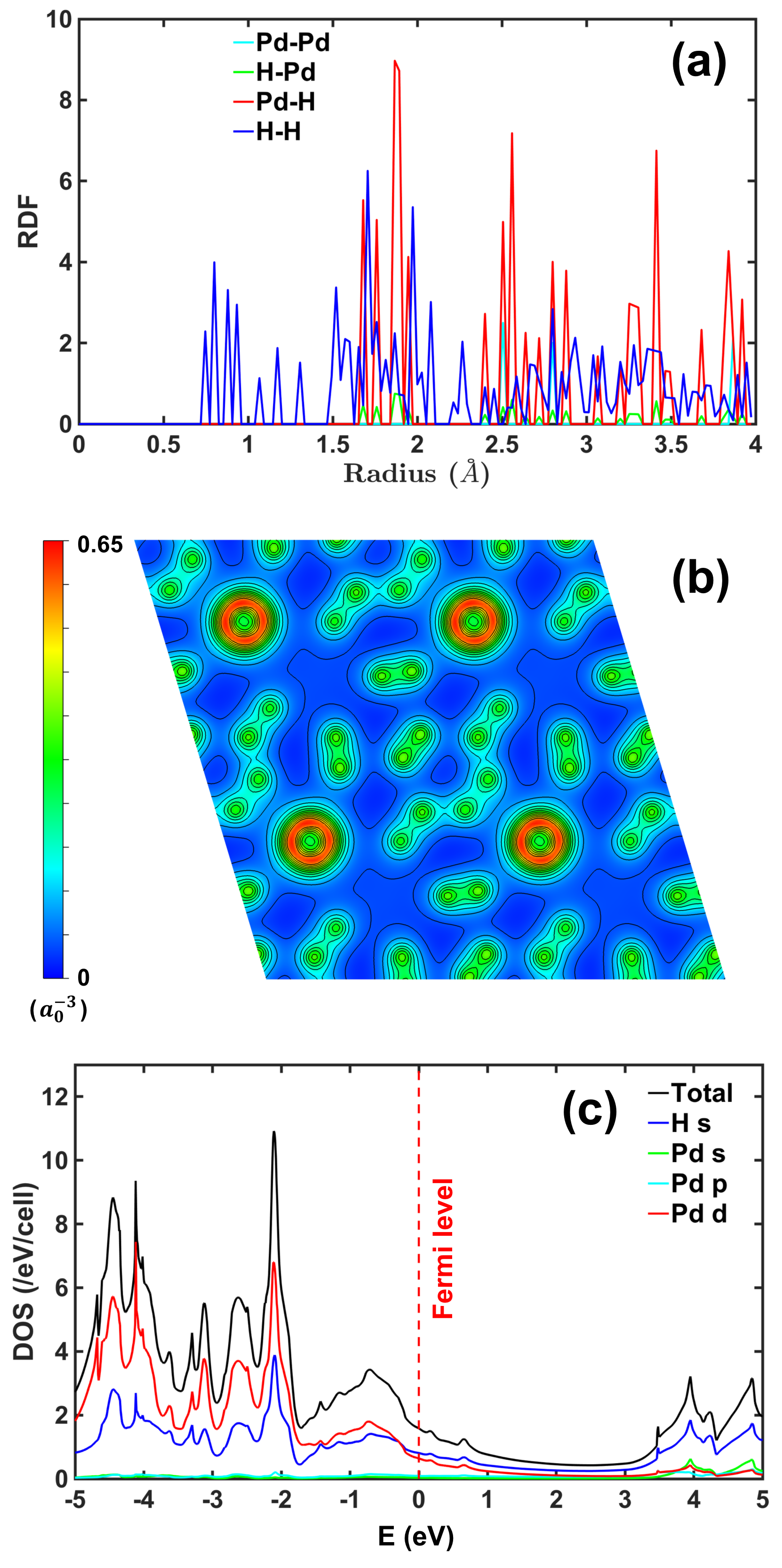}
\centering
\caption{Properties of $Cmcm$ \ce{PdH12} at 150 GPa: (a) Radial distribution function (RDF). The predominant bonding is H-H and Pd-H. Note that the minimum H-H distance is about 0.74 \AA, close to H-H bond in molecular \ce{H2}. (b) Charge density distribution within the (001) plane. The charge density between H and H is higher than that between Pd and H. The unit of the charge density is $a_0^{-3}$, where the Bohr radius $a_0$ = 0.529 {\AA}. The contour lines are spaced by 0.05 $a_0^{-3}$. (c) Density of states (DOS) and partial density of states (PDOS). The red dashed line is the Fermi level.}
\label{fig:PdH12-properties}
\end{figure}

The most probable stable structure found for \ce{PdH12} at 150 GPa is monoclinic with $Cmcm$ space group (Fig. \ref{fig:0&150-PdHn}). This structure is distinct from those predicted as thermodynamically stable phases for rare-earth superhydrides, which are based on clathrate or cage-like structures\cite{liu2017potential,peng2017hydrogen}. On the other hand, the structure has similarities to those predicted for \ce{MgH12} and \ce{MgH16}, which consists of molecular \ce{H2} units having a range of nearest neighbor distances\cite{PhysRevB.87.054107}. The $Cmcm$ structure of \ce{PdH12} consists of a 3D network of Pd-centered H polyhedra bridged by H-H covalent bonds with the space in-between filled with \ce{H2} molecules. Interestingly, viewing perpendicular to the monoclinic c axis, all the H atoms are arranged in 2D layers, which are stacked together along the c axis. Fig. \ref{fig:PdH12-properties}(a) shows its radial distribution function (RDF). The nearest neighbor Pd-H distances span 1.7-1.9 \AA, 
giving an effective coordination of Pd by H of 14. On the other hand, the H-H distances span a wide range and consist of two groups (1) 0.7-1.0 \AA, corresponding to \ce{H2} molecules between polyhedra and the bridging \ce{H2} units that connect polyhedra, and (2) 1.5-2.0 \AA, corresponding to neighboring corner H atoms within a polyhedron. The charge density distribution of a (001) plane is plotted in Fig. \ref{fig:PdH12-properties}(b). The charge density between H-H is higher than that between Pd-H, indicating that covalent H-H bonds are dominant even under pressures of 150 GPa. This is likely due to zero pressure electronegativities of Pd and H being very close. We conclude that this equivalence persists to high compressions, as evident by the lack of significant charge transfer even under these extreme conditions. Nevertheless, bonding between Pd and H is apparent from the charge density map. The electronic density of states (DOS) is shown in Fig. \ref{fig:PdH12-properties}(c), where the total DOS is decomposed to contributions from different orbitals, H-s, Pd-s, Pd-p and Pd-d. The considerable DOS at the Fermi level indicates that the monoclinic \ce{PdH12} is a metal under these conditions. The dominant partial DOS are from H-s and Pd-d showing strong hybridization, both contributing to conducting electrons. 

The H-H distances for the rare-earth superhydrides are in the range of $\sim$1.1 \AA~ at high pressures\cite{liu2017potential,peng2017hydrogen}. Recently, evidence for H-H distances below 1.6 \AA~ was found in \ce{ZrV2Hx}, even at ambient pressure, from inelastic neutron scattering\cite{borgschulte2020inelastic}, in violation of the so-called Switendick\cite{switendick1979band} criteria for the minimum H-H distances of 2.1 \AA~ in common hydrides. In order to analyze how the H-H distance changes with composition and pressure, the distribution of minimum H-H distances of the low enthalpy structures at both zero pressure and 150 GPa is plotted in Fig. S6. The composition has a large influence on the minimum H-H distance. At zero pressure, the minimum H-H distance is around 2.2-3.1 \AA~ for PdH, but drastically drops to about 0.8 \AA~ for \ce{PdH2}, close to the H-H distance in \ce{H2}, 0.74 \AA. Further increasing H content does not cause significant changes, and the minimum H-H distance is pinned by the H-H distance in \ce{H2} molecule. A similar trend holds for 150 GPa, though the minimum H-H distance usually shrinks compared with that at zero pressure, due to large compression. However, the H-H distance is close to that of free \ce{H2}.

Having identified the possible stable structures, we assess the phase stability from an enthalpy convex hull analysis.  Using an ensemble of functionals, we generate an ensemble of convex hulls, which are shown in Figs. S7 and S8.  Unsurprisingly, We  find that no Pd superhydrides are thermodynamically stable at zero pressure. At 150 GPa, there are still no thermodynamically stable superhydrides although \ce{PdH12} is only slightly unstable, and the stability of superhydrides is greatly increased compared to zero pressure.

\section*{Moderate-pressure electrochemical synthesis}

Superhydrides were first documented experimentally using megabar high pressure diamond-anvil cell laser-heating techniques, leading to the discovery of near-room-temperature superconductivity in \ce{LaH10}\cite{geballe2018synthesis,PhysRevLett.122.027001,hemley2019road}. This result was subsequently confirmed\cite{Drozdov2019}, and other metal superhydrides have subsequently been observed\cite{Zurek2019}. We now discuss an alternate approach, \psquared, to synthesizing dense metal superhydrides by combining pressure and electrochemistry (i.e. electrode potential). In the \psquared~ approach, an electrode consisting of a metal (or conducting metal hydride) is loaded with hydrogen by holding at an appropriate electrode potential using an electrolyte consisting of mobile protons.  The proton-conducting membrane could be an aqueous electrolyte solution, polymer electrolyte membrane (e.g., Nafion)\cite{benck2019producing}, proton-conducting ceramic electrolytes\cite{benck2019producing}, and solid acid proton conductors\cite{haile2007solid}. The electrolytes provide a way to tune the activity of mobile protons and kinetics of reactions at electrode-electrolyte interfaces.

Continuing with the example of the Pd-H system, the hydrogen loading reaction for the metal electrode and an electrolyte containing mobile protons is given by:
\begin{equation}
    \mathrm{Pd + n(H^+ + e^-) \Longleftrightarrow PdH_n}
\end{equation}
with the associated Gibbs Free Energy of the reaction:
\begin{equation}
    \mathrm{\Delta G = G_{PdH_n} -  G_{Pd} - n G_{H^+} - n G_{e^-}}.
\end{equation}
The free energy of protons at unit activity and electrons at electrode potential zero on the standard hydrogen electrode scale can be related to the free energy of hydrogen gas.  Thermodynamic corrections can then be added to account for the effect of electrode potential and activity of protons\cite{norskov2004origin}. The computational hydrogen electrode equation provides the relation, $\mathrm{G_{H^+ (a_H^+ = 1)} + G_{e^- (U = 0)} = \frac{1}{2} G_{H_2}}$.  This provides the relation, 
\begin{equation}
    \mathrm{\Delta G = G_{PdH_n} -  G_{Pd} - \frac{n}{2} G_{H_2} + neU_{SHE} -n k_BT ln(a_{H^+})}.
\end{equation}
This relation allows us to construct an electrochemical phase diagram for loading hydrogen into a material as a function of pH and electrode potential.   Lowering the electrode potential (making it more negative) or increasing the activity of protons enables loading higher amounts of hydrogen.

However, in a practical device, at negative potentials, metals tend to catalyze the hydrogen evolution reaction\cite{norskov2005trends}, given by,
\begin{equation}
    \mathrm{2H^+ + 2e^- \Longleftrightarrow H_2}.
\end{equation}
Hence, electrochemical loading needs to compete with the hydrogen evolution reaction.  Electrolyte formulations can suppress the hydrogen evolution reaction kinetically through superconcentrated electrolytes\cite{suo2015water} or other suppressing mechanisms.  However, we set the limit for electrochemical synthesizability at the potential where reaction free energy for hydrogen evolution on the catalyst surface is thermodynamically downhill, which is determined by the free energy of adsorbed hydrogen on the palladium surface\cite{norskov2005trends}.  While the \psquared~ approach is demonstrated with a proton conductor, a similar scheme can be constructed with hydride ion conductors\cite{Verbraeken2014} (for e.g. $\mathrm{Pd + nH^- \Longleftrightarrow PdH_n + ne^-}$).

\begin{figure}
\includegraphics[width=0.62\linewidth]{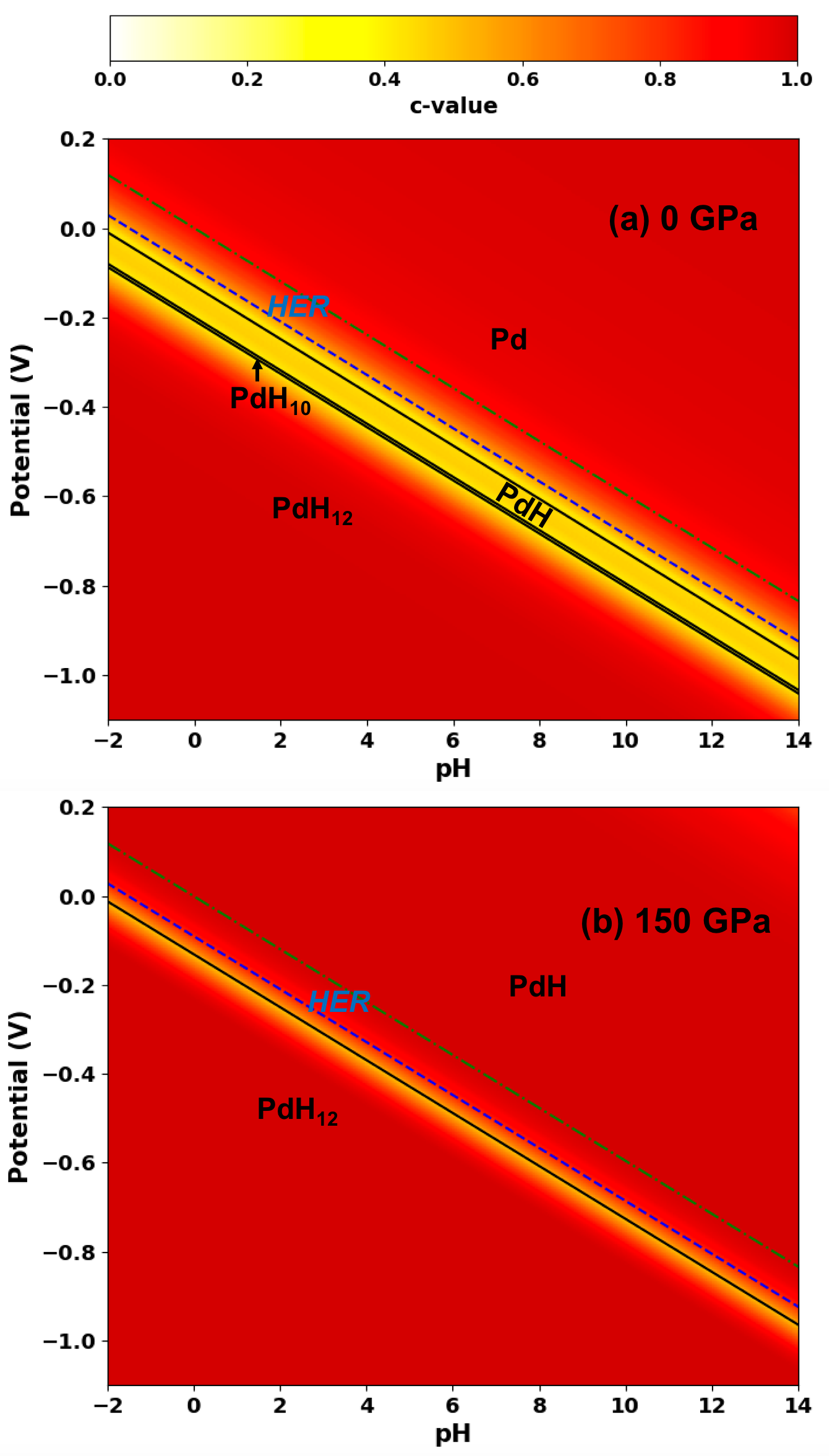}
\centering
\caption{Pourbaix diagram by BEEF-vdW ensmebles at (a) zero pressure and (b) 150 GPa. The solid black lines are phase boundaries calculated by the best-fit BEEF-vdW functional. The dotted dashed green line represents the equilibrium HER (hydrogen evoluation reaction) on Pd, while the blue dashed line takes overpotential into consideration.}
\label{fig:pb-uq}
\end{figure}

The electrochemical phase diagram incorporating uncertainty analysis at ambient pressure is shown in \cref{fig:pb-uq}.  At ambient pressure, at $pH = 0, ~(a_{H^+} = 1)$, we find that electrochemical loading of even the PdH phase is challenging and will compete with the hydrogen evolution reaction, as observed experimentally\cite{benck2019producing}. As Pd catalyzes hydrogen evolution with almost no overpotential\cite{norskov2005trends}, thus suppressing hydrogen evolution is the only approach to accessing these phases at ambient conditions.  Further, it is likely that the bulk loading reaction will have slower kinetics than surface catalyzed hydrogen evolution reaction making this even more challenging.  Next, we explore the effect of very high pressures, and find that it is possible to produce \ce{PdH12} below the hydrogen evolution potentials.  

%Moderate-pressure electrochemistry has been little explored, but there have been some attempts at modest pressures.\cite{giovanelli2004electrochemistry,havens2009high}   Here, we combine both high-pressure and electrochemistry to understand the electrochemical phase diagram at 150 GPa \pg{compatibility with previous moderate pressure sentence?}, which is shown in Fig. \ref{fig:pb-uq}.  Interestingly, we now find that it is possible to load several superhydride phases \pg{only PdH12?} around the potentials for hydrogen evolution reaction.

\begin{figure}
\includegraphics[width=\linewidth]{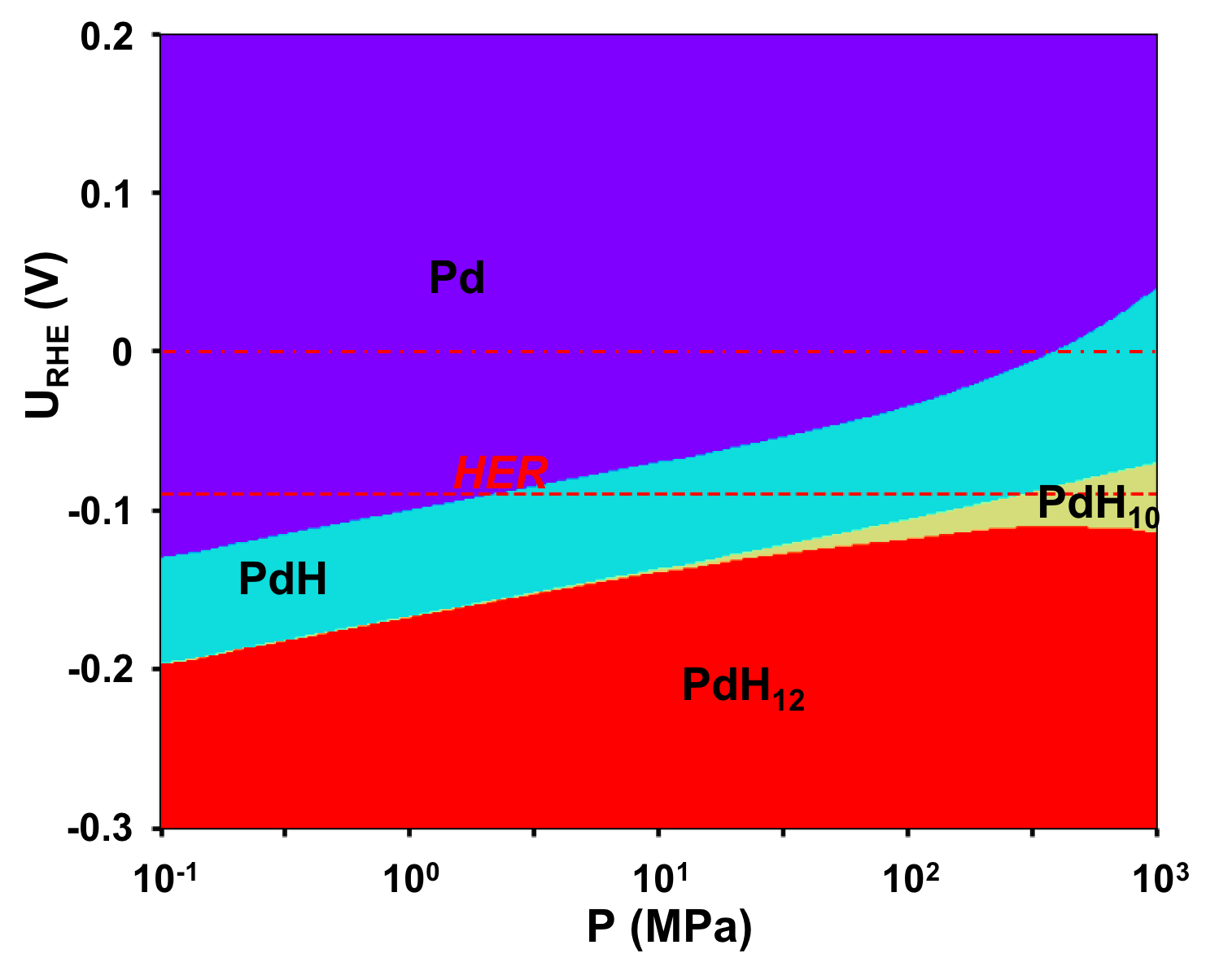}
\centering
\caption{Pressure dependent Pourbaix diagram of Pd-H by the best-fit BEEF-vdW functional at 300 K. The dotted dashed red line represents the equilibrium HER (hydrogen evolution reaction) on Pd, while the red dashed line takes overpotential into consideration.}
\label{fig:U-P}
\end{figure}

The above analysis raises the intriguing question of whether such palladium superhydrides could be synthesized at modest pressures, e.g., at 100 MPa (kilobar) versus 100 GPa (megabar) conditions. Although electrochemical studies have been performed over the years in the 100 MPa range (maximum of ~1 GPa\cite{doi:10.1021/j100203a057,giovanelli2004electrochemistry}), the field remains largely unexplored. To examine the possibility to stablize Pd superhydrides at modest pressures, we calculate a comprehensive pressure dependent Pourbaix diagram at 300 K, with the reversible hydrogen electrode (RHE) as the reference (Fig. \ref{fig:U-P}). At decreasing potentials, the phase transition sequence is $\ce{Pd} \rightarrow \ce{PdH} \rightarrow \ce{PdH10} \rightarrow \ce{PdH12}$. The most thermodynamically accessible Pd superhydride, \ce{PdH10} has a very narrow potential window near ambient pressure, which is gradually enlarged with increasing pressure. The phase boundary between PdH and \ce{PdH10}, can be fitted as the following relation: 
\begin{equation}
    \mathrm{U=-0.168+0.0297*log_{10}(P)+8.83*10^{-5}\times P^{2/3}},
\end{equation}
where U is the electrode potential on the RHE scale (in V) needed to transform PdH to \ce{PdH10} under a given pressure P (in MPa).   The above relation has good linearity between U and logP near the ambient pressure.  The power of the electrochemical driving force is illustrated by the fact that an order of magnitude reduction of the transition pressure can be achieved by only 0.03 V change in the electrode potential. When the pressure approaches $10^3$ MPa (1 GPa), non-linearity of the relation becomes significant. The superhydride \ce{PdH10} can be stablized at about 300 MPa (0.3 GPa) at the onset potential for hydrogen evolution. Using superconcentrated electrolytes that can suppress HER further, it is expected that \ce{PdH10} can be stablized under even lower pressure operating at a more negative potential.  

\begin{figure}
\includegraphics[width=\linewidth]{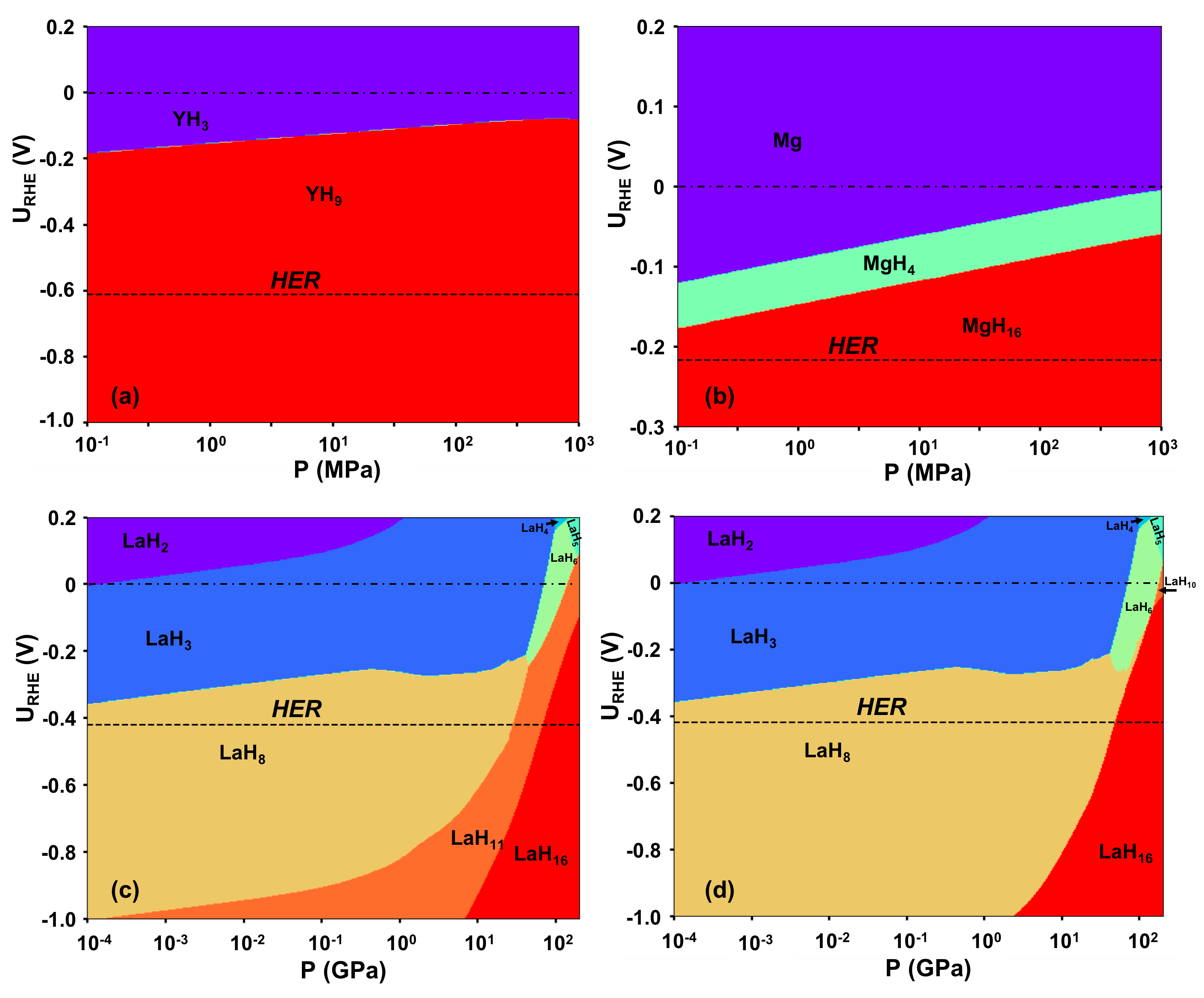}
\centering
\caption{Pressure dependent Pourbaix diagram by the best-fit BEEF-vdW functional at 300 K: (a) Y-H, (b) Mg-H, and La-H (c) with \ce{LaH11} and (d) without \ce{LaH11} considered. The dotted dashed lines represent the equilibrium HER (hydrogen evolution reaction) on the metal, while the dashed line takes overpotential into consideration.}
\label{fig:M-H}
\end{figure}

\section*{Other superhydrides}

From an electrochemical synthesis standpoint, palladium is among the most difficult as it catalyzes hydrogen evolution with negligible overpotential\cite{norskov2005trends}.  To demonstrate the generality of this strategy, the pressure dependent Pourbaix diagrams were calculated for three other hydride systems, Y-H, Mg-H and La-H (Fig. \ref{fig:M-H}), all of which have been predicted or observed to have high-temperature superconductors. For all these systems, calculations similar to those above for Pd-H indicate that \ce{YH9}, \ce{MgH4}, \ce{MgH16} and \ce{LaH8} can be synthesized under ambient pressure at potentials without HER. As for Pd-H, the critical potential becomes less negative with increasing pressure. The results for these different metal-hydrogen systems thus demonstrate that using electrochemistry to drastically lower the synthesis pressure of superhydrides is a general approach. 

Most notable are the results for La-H, and the near-room-temperature superconducting phase \ce{LaH10}\cite{PhysRevLett.122.027001,hemley2019road,Drozdov2019}. The calculated stability to 200 GPa for La-H shows a rich set of superhydrides, the stabilizing pressures of which can be significantly reduced (in tens of GPa) by the electrode potential above HER. For instance, \ce{LaH16} can be stabilized below 100 GPa at the HER potential while an extreme pressure above 200 GPa is needed without an electrochemical driving force. It is also noted that the experimentally verified \ce{LaH10} phase is stabilized only after removing \ce{LaH11} in the underlying data, which may be due to the errors caused by neglecting higher order corrections such as anharmonicity \cite{liu2018} and nuclear quantum effects \cite{Errea2020} in the calculations. Such corrections usually enhances the stability of the high density structures to lower pressures \cite{liu2018}.   More interestingly, the potential not only decreases the critical pressure, but also stabilizes some phases which cannot be stabilized regardless of pressure at zero potential, such as \ce{LaH4}, \ce{LaH5} and \ce{LaH8}. Such a new degree of freedom therefore brings a broad new world that is largely unexplored and promisingly fertile. 

\section*{Conclusions}
%In summary, we find a remarkable result that novel hydrides of palladium including palladium superhydrides, can be synthesized by combining electrochemical and high pressure techniques. Moreover, the results show that new phases are accessible at even modest pressures of 300 MPa, which is readily accessible using existing high pressure methods. Given that palladium is among the most active metals for hydrogen evolution, we believe that it will be even easier to synthesize other metal superhydrides.  This work should serve to open up the new frontier of high-pressure electrochemistry to produce exciting new materials with broad applications\cite{berlinguette2019revisiting,FloresLivas2020}.
%Using first-principles density functional theory calculations and particle swarm structure search, the phase stability of superhydrides of palladium have been studied. From an analysis of the electrochemical loading of hydrogen in Pd under a broad range of pressures, remarkably we find that \ce{PdH10} can be electrochemically synthesized at pressures below 300 MPa before the onset of hydrogen evolution. Combining pressure and electrode potential thus offers an alternate route to synthesize metal superhydrides and other novel materials at currently accessible static pressure conditions, and a framework for such synthesis has been provided. Numerous extensions of existing electrochemical and high pressure techniques could lead altogether new materials created under an even broader range of pressures. 
Using first-principles density functional theory and particle swarm structure search calculations, the phase stability of metal superhydrides have been studied. Focusing on the Pd-H system, we find from an analysis of the electrochemical loading of hydrogen in the metal under a broad range of pressures, that \ce{PdH10} can be electrochemically synthesized before the onset of hydrogen evolution. Remarkably this is predicted to occur at even modest pressures of about 300 MPa, which is readily accessible using existing high pressure methods. Given that palladium is among the most active metals for hydrogen evolution, we suspected significant effects of electrochemical loading on the synthesis of other hydrides.  Indeed, we demonstrate the generalizability of this pressure-potential (\psquared~) approach for La-H, Y-H and Mg-H, often yielding 10-100 times reduction in pressure needed for stabilizing a particular phase, as well as stabilizing new phases that cannot be done purely by either pressure or potential. Combining pressure and electrode potential thus offers an alternate route to synthesize metal superhydrides and other novel materials at currently accessible static pressure conditions, and a framework for such synthesis has been provided. Numerous extensions of existing electrochemical and high pressure techniques could lead altogether new materials created under an even broader range of pressures. This work should serve to open up the new frontier of high-pressure electrochemistry to produce exciting new materials with broad applications\cite{berlinguette2019revisiting,FloresLivas2020}.

% \begin{figure}
% \includegraphics[width=\linewidth]{pourbaix-PBE-150GPa.png}
% \centering
% \caption{Pourbaix diagram by PBE at 150 GPa. The red dashed line represents HER on Pd, which is within the stability region of PdH. Compared with 0 GPa, it can be seen that \ce{PdH12} is more achievable under electrochemical environments.}
% \label{fig:pb-pbe-150}
% \end{figure}

\begin{methods}
\subsection{Calculation Details:} All DFT calculations involving the BEEF-vdW exchange correlation functional were run using GPAW software using the Atomic Simulation Environment (ASE) \cite{PhysRevB.71.035109,Enkovaara_2010,ase-paper}. A real space grid with spacing of 0.16 \AA ~ is used for the representation of electronic wavefunctions, and a k-point density of larger than 30 \AA~  in reciprocal space was used in each dimension. For each material, the geometry is relaxed to a maximum force of less than 0.01 eV/\AA.

\subsection{Structure Search:} The particle swarm optimization is employed for structure search, using the CALYPSO code\cite{Wang2010,Wang2012}. Since only the superhydrides are of interests here, the compositions PdH$_n$ (n is an integer and $1 \leq n \leq 12$) and \ce{Pd3H4} which is the highest Pd hydride reported in experiments so far, totaling 13 compositions in all. For each composition, one unit cell is allowed to have 1-4 formulas. For a fixed number of formula at a given composition, about 1000 structures were searched during the structure evolution.

\subsection{Bayesian Error Estimation Funtional:} For each material, a collection of functionals at the level of the generalized gradient approximation (GGA) were used as described below. Error estimation was carried out using the Bayesian error estimation functional with van der Waals correction \cite{BEEF}. This empirically fit functional generates an ensemble of functionals that are small perturbations away from the best fit functional in exchange correlation space.

The exchange-correlation energy for the BEEF-vdW is given in Ref. \citenum{BEEF} as
\begin{equation}
    E_{xc}=\sum_{m} a_m \int \epsilon_x^{UEG}(n)B_m[t(s)]d\mathbf{r} 
	+\alpha_c E^{LDA-c}+(1-\alpha_c)E^{PBE-c}+E^{nl-c}.
\end{equation}
Here $B_m$ is the m$^{th}$ Legendre basis function, each of which has a corresponding expansion coefficient $a_m$. The expansion coefficients, as well as the $\alpha_c$ parameter that mixes the local density approximation (LDA) and PBE\cite{PhysRevLett.77.3865} exchange correlation functionals, have been pre-fit with respect to a range of data sets as described in Ref. \citenum{BEEF}. Additionally within the functional is the $E^{nl-c}$ non-local correlation term implemented via the vdW-DF2\cite{lee2010higher} method. The method to generate the ensemble of functionals was tuned such that the spread of the predictions of the functionals matches the error of the main self consistent functional with respect to the training and experimental data on which it was originally trained. Each of these functionals can then provide a non self consistent prediction of energy and therefore allows for a computationally efficient yet systematic way of understanding the sensitivity of the final prediction with respect to small changes in exchange correlation functional. 

\subsection{Confidence-value (c-value) Calculations:} We use the confidence value (c-value)\cite{houchins}, for determining the uncertainty associated with the choice of the functional.  We define c-value to be the fraction of the ensemble that predict a certain structure to be the ground state.  For a fixed composition, this simply involves counting the fraction of functionals that predict a particular structure to be the ground state.  This framework can be expanded to construct a c-value associated with a Pourbaix diagram.  In this case, the c$(U,pH)$, for a specific phase is defined as the fraction of functionals that predict it to have the lowest free energy at a given potential and pH, given by
\begin{equation}
c_i(U,pH) =  \frac{1}{N_{ens}} \sum_{n=1}^{N_{ens}} \prod_{j\neq i} \Theta(\Delta G_{j}^n(U,pH)-\Delta G_{i}^n(U,pH))
\end{equation}
\noindent where the summation is over number of ensembles, $\mathrm{N_{ens}} = 2000$ here, and the product is over all the remaining possible phases. $\Theta(x)$ denotes the Heaviside step function.  At any given U and pH, $i, j \in S$, the set of all considered phases.

\subsection{Topological analysis:} The coordination number is determined using the CrystalNN class based on a Voronoi algorithm in pymatgen\cite{Ong2013}. The framework of the crystal structure and its dimensionality are identified using the Zeo++ code based on the Voronoi decomposition\cite{Willems2012}, where radii of 0.5 {\AA} and 1.6 {\AA} are adopted for H and Pd respectively.
\end{methods}

\begin{addendum}
\item This work was partially supported by Google (P. G. and V. V.) and by the DOE/NNSA and NSF-DMR (R.J.H.). The authors thank Hanyu Liu and Yanming Ma for sharing structure files of the Y-H, Mg-H and La-H systems. The Authors would like to thank helpful discussions with Yet-Ming Chiang, Matt Trevithick and Florian Metzler.
\item[Contributions] P.G. and V.V. conceived the idea for the project, and R.J.H. provided input on the high pressure calculations and experiments. P.G. ran all the DFT calculations. All authors discussed the results and jointly wrote the manuscript.
\item[Competing Interests] V.V. and P.G. are inventors on a provisional patent application, 63/028,265 related to electrochemical synthesis of metal superhydrides.
\item[Correspondence] Correspondence and requests for materials should be addressed to V. Viswanathan (email: venkvis@cmu.edu) and R.J.Hemley (rhemley@uic.edu).

\item[Additional information] 
Supplementary Information attached and consists of Supplementary Figures 1-11.

\item[Data availability] 
The data that support the findings of this study will be made publicly available on Github.

\item[Code availability] 
The custom code for calculating and plotting the phase diagrams based on first-principles data presented in the paper will be made publicly available on Github.
\end{addendum}

\clearpage

\bibliography{cite}

\includepdf[pages=-]{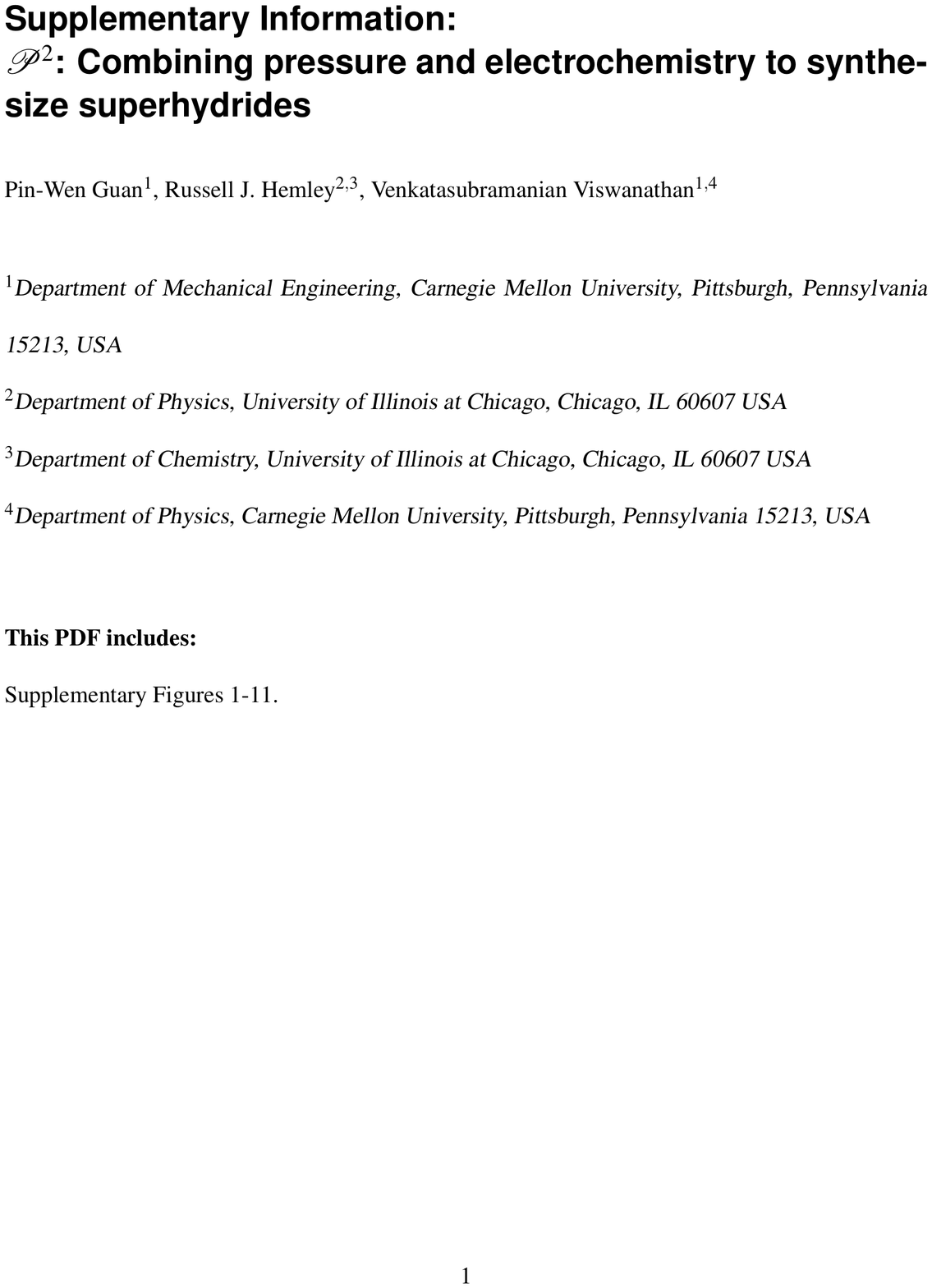}

\end{document}